\documentclass[aps,prl,twocolumn,showpacs]{revtex4}
\usepackage{graphicx}
\usepackage{amsmath}

\begin{document}

\title{Chirality and Orbital Order in Charge Density Waves}

\author{Jasper van Wezel}
  \affiliation{Materials Science Division, Argonne National Laboratory, Argonne, IL 60439, USA}

\begin{abstract}
We show that the recently observed chirality in the charge ordered phase of TiSe$_2$ can be understood as a form of orbital ordering. The microscopic mechanism driving the transition between the novel chiral state and the non-chiral charge density wave is discussed, and shown to be of a general form, thus allowing for a broad class of materials to display this type of orbitally ordered chiral charge density wave.
\end{abstract}


\maketitle

{\it Introduction.}--It has recently been suggested that the layered, quasi two-dimensional charge density wave (CDW) compound 1T-TiSe$_2$ may possess a chiral charge ordered phase \cite{Ishioka:2010ez,JvW:Physics11}. That is, the dominant propagation vector among the three components which make up the triple-$q$ CDW phase of TiSe$_2$, rotates as one progresses from one atomic layer to the next. Although chirality is a quite common property of spin density waves \cite{Overhauser:1962gc}, TiSe$_2$ so far seems to stand alone as a material in which the formation of a charge density wave is accompanied by the emergence of a helical symmetry. The reason that spin ordered states are so much more prone to being chiral is due to the vector nature of their order parameter: rotating one vector while propagating along another trivially yields a helical pattern. For the purely charge ordered state on the other hand, the order parameter is a scalar quantity, and it is not immediately obvious how it may give rise to a chiral state.

It is well known that the inversion symmetry of an atomic lattice may be spontaneously broken in charge ordered materials by the ionic displacements associated with the ordering. The combination of a commensurate CDW and a lattice without inversion symmetry gives rise to the emergence of ferroelectricity in materials like SnTe \cite{Littlewood:1983fg}. For a sliding, incommensurate CDW on the other hand, the breakdown of inversion symmetry in the lattice may lead to dramatic hysteresis effects such as that seen in NbSe$_3$ \cite{1984PhRvB..30.3530M,Gruner:1988wl}. The measurements of Ishioka et al. have now shown that it is also possible for a CDW to break inversion symmetry with an axial vector instead of a polar one, resulting in a chiral state \cite{Ishioka:2010ez,JvW:Physics11}. 

In this paper, we will discuss the microscopic mechanism which underlies the formation of a chiral CDW in TiSe$_2$, and point out that it is closely linked to the existence of orbital order in the chiral state. We will show that this type of combined orbital and charge order may in fact be expected to be a generic property of a broad class of CDW materials and discuss the prerequisites for finding chiral charge order in other materials.

{\it Polarization.}--Although TiSe$_2$ is the only known compound in which chirality arises from a charge density wave transition, there are other materials which have a chiral crystal structure. Of these, elemental Tellurium and Selenium stand out, because it has long been suggested that their chirality may be understood as originating from a charge density wave in a (non-existent) cubic parent phase \cite{Fukutome:1984vs}. In such a cubical structure, the two-third filled $p$-shell results in three bands crossing the Fermi energy, each dominated by a differently oriented $p$-orbital. These orbitals, to a first approximation, form non-interacting one-dimensional chains, so that their Fermi surfaces are well nested, and each of the three bands undergoes a Peierls transition with the same nesting vector $Q$. 

\begin{figure}[t]
\centerline{{\includegraphics[width=0.6 \columnwidth]{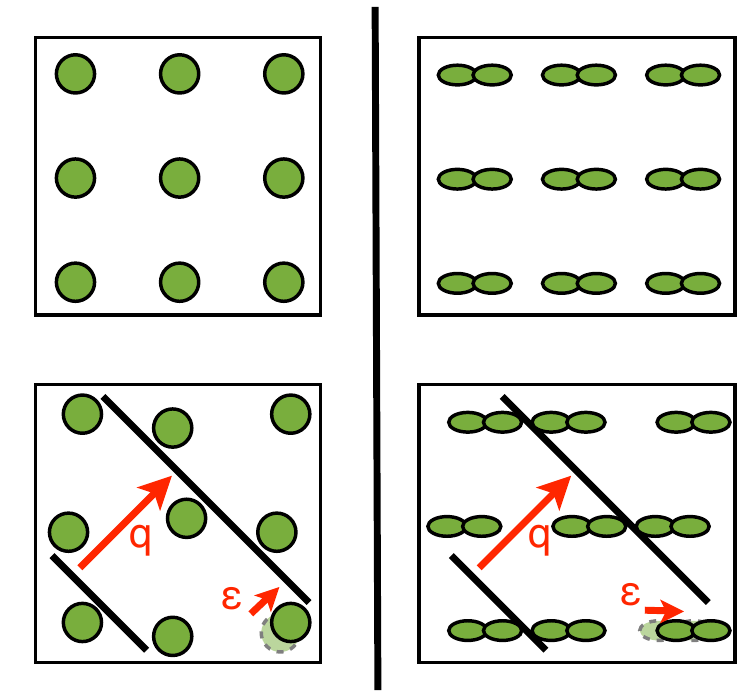}}}
\caption{(Color online) Schematic picture of the displacement wave $\vec{u}=u_0 \hat{\epsilon} \cos(\vec{q}\cdot\vec{x})$. Left: in a purely isotropic system the lattice distortions move the ions directly towards the charge maxima, and the strain wave is purely longitudinal. Right: if the lattice consists of anisotropic orbitals, the strain wave may acquire transversal components due to the electron-phonon coupling matrix elements.}
\label{polarizations}
\end{figure}

Because the electron-phonon coupling is finite in any real material, the electronic Peierls instability is always accompanied by a corresponding periodic displacement of the ions in the underlying lattice. This periodic displacement wave (PDW) follows the redistribution of electronic charge in order to minimize Coulomb energy. The ability of the lattice distortions to align with the charge modulation however, is limited by the anisotropy of the electron-phonon coupling:
\begin{align}
u_i \propto \sum_j \eta_{ij} \frac{\partial}{\partial x_j} \alpha,
\end{align}
where the lattice distortion $\vec{u}(\vec{x})$ is related to the charge modulation $\alpha(\vec{x})$ through the electron-phonon coupling matrix elements $\eta_{ij}$. The anisotropy of the $p$-orbitals in Te and Se leads to an anisotropic electron-phonon coupling for each orbital sector, which results in a partly transversal polarization of the displacement wave, as shown in Fig. \ref{polarizations}. Combining three such polarized PDW --one for each orbital orientation-- with the correct relative phase differences can be seen to result in the chiral lattice structure observed for elemental Te and Se \cite{Fukutome:1984vs,JvW:Physics11, tbp}.
The central role played by the vector nature of the PDW in Te and Se suggests that also the occurrence of the chiral phase in TiSe$_2$ may be connected to the polarization directions of its displacement waves. 

\begin{figure}[t]
\centerline{{\includegraphics[width=0.95 \columnwidth]{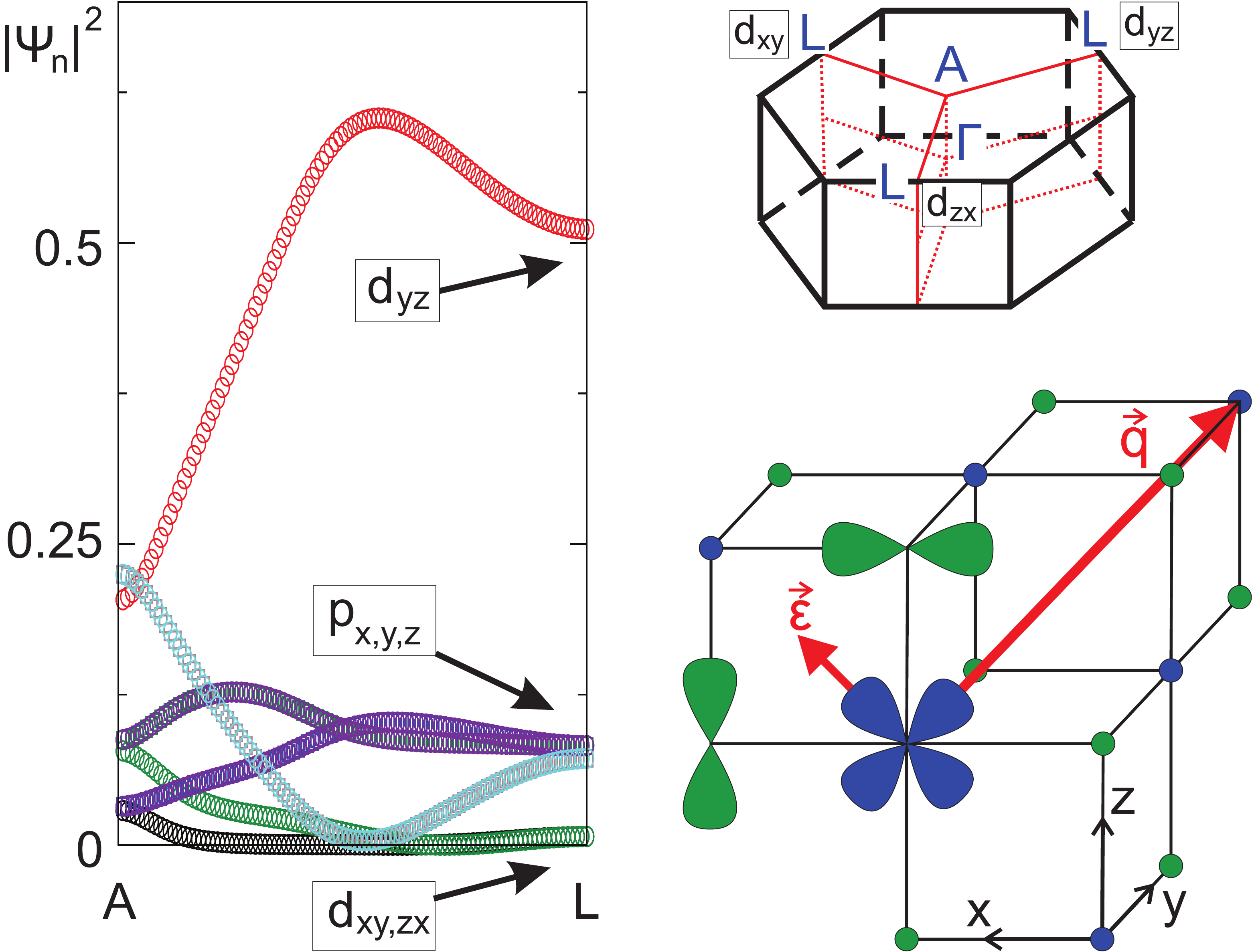}}}
\caption{(Color online) Left: The orbital character of the lowest electron band along the $A$-$L$ high symmetry direction. Right top: schematic representation of the first Brillouin zone, including the dominant orbital character at each of the $L$ points. Right bottom: One layer of TiSe$_2$, showing the orbitals involved in the charge transfer along one of the $\Gamma$-$L$ directions. Shown also are the CDW propagation vector corresponding to this particular $\Gamma$-$L$ direction, and the resulting ionic displacement. Notice that the coordinate system used here is not the standard one for a hexagonal layered system. The orbitals $d_{xy}$, $d_{yz}$ and $d_{zx}$ are related to each other by rotations of $2 \pi /3$ around the crystallographic $c$-axis (the $(\bar{1} \bar{1} 1)$ direction).}
\label{bands}
\end{figure}

Although the mechanism driving the CDW formation in TiSe$_2$ remains uncertain even after decades of study \cite{1977SSCom..22..551W, Hughes:1977tg, Whangbo:1992vu, Cercellier:2007tr, vanWezel:2010uo, vanWezel:2010uc, Calandra:2011bo, Rossnagel:2011jj}, it is clear that the transition is accompanied by a transfer of electronic spectral weight from the maximum of the Se $4p$ band at the centre of the first Brillouin zone to the three minima of the Ti $3d$ bands at the zone boundaries \cite{Cercellier:2007tr}. Using a tight-binding fit to the known band structure of TiSe$_2$ \cite{vanWezel:2010vq}, we can determine the orbital character of these bands. As is shown in Fig. \ref{bands}, the electrons at the $L$ point come predominantly from a single type of Ti $t_{2g}$ orbital. 

Looking at the other two inequivalent $L$ points, we see that the orbital character changes as we go around the Brillouin zone (see Fig. \ref{bands}). The same picture holds for the hole states at the $\Gamma$ point, which are dominated by different Se $4p$ orbitals at each side. Thus the three propagation vectors of the CDW in TiSe$_2$ (the three inequivalent $\Gamma$-$L$ vectors) cause charge transfer in three distinct orbital sectors, in close analogy to the situation in elemental Tellurium. In fact, as shown in Fig. \ref{bands}, the relative orientations of the $t_{2g}$ and $p$-orbitals forces the displacement waves in TiSe$_2$ to be purely transversal. The PDW in TiSe$_2$ may thus be written as the sum of three components $\vec{u}_i=u_0 \hat{\epsilon_i} \cos(\vec{q}_i \cdot \vec{x} + \varphi_i)$, with three different sets of propagation vectors and polarizations. The values of the relative phases $\varphi_i$ remain to be determined.
%

{\it Relative Phases.}--We construct a Ginzburg-Landau theory to describe the microscopic interactions between the electronic density modulations and the lattice, which determine the relative phases of the PDW components. The order parameter consists of the modulation $\alpha(\vec{x})$ of the average charge density, which can be written as a sum of three complex components $\psi_j= \psi_0 e^{i\vec{q}_j\cdot \vec{x} + \varphi_j}$. If we assume  the amplitudes of all components to be equal and fix the propagation vectors at their preferred values, the free energy in its most general form can be written as \cite{McMillan:1975wg}:
\begin{align}
F &= \int d\vec{x} \left\{ \ a\alpha^2 + b \alpha^3 + c \alpha^4 \right. \notag \\
&+ \left. d \left[ |\psi_1 \psi_2 |^2 + |\psi_2 \psi_3 |^2 + |\psi_3 \psi_1 |^2 \right] \right\}.
\label{F}
\end{align}
The cross terms in the last line signify the competition between the various CDW components over the available Fermi surface, which determines whether the structure will be of the single-$q$ or triple-$q$ form \cite{McMillan:1975wg}. 

For the evaluation of the integrals it is essential to take into account the Umklapp processes associated with the presence of a discrete lattice. This can be done by forcing the spatially varying part of the coefficients to reflect the symmetry of the lattice, as well as the internal structure of the unit cell \cite{McMillan:1975wg}:
\begin{align}
a = a_0 + a_1 \sum_i e^{i\vec{G}_i \cdot \vec{x}} \left( 1 + \gamma e^{i\vec{G}_i \cdot \vec{R}_{\text{Se1}}} + \gamma e^{i\vec{G}_i \cdot \vec{R}_{\text{Se2}}} \right) + .. \notag
\end{align}
Here $\vec{G}_i$ are the shortest reciprocal lattice vectors and $\vec{R}_{\text{Se1,2}}$ denote the positons of the two Se atoms within the unit cell. The factor $\gamma$ reflects the difference between the electron-phonon couplings on the Ti and Se sites. Higher order terms in the expansion include longer reciprocal lattice vectors. 
\begin{figure}[t]
\centerline{{\includegraphics[width=0.8 \columnwidth]{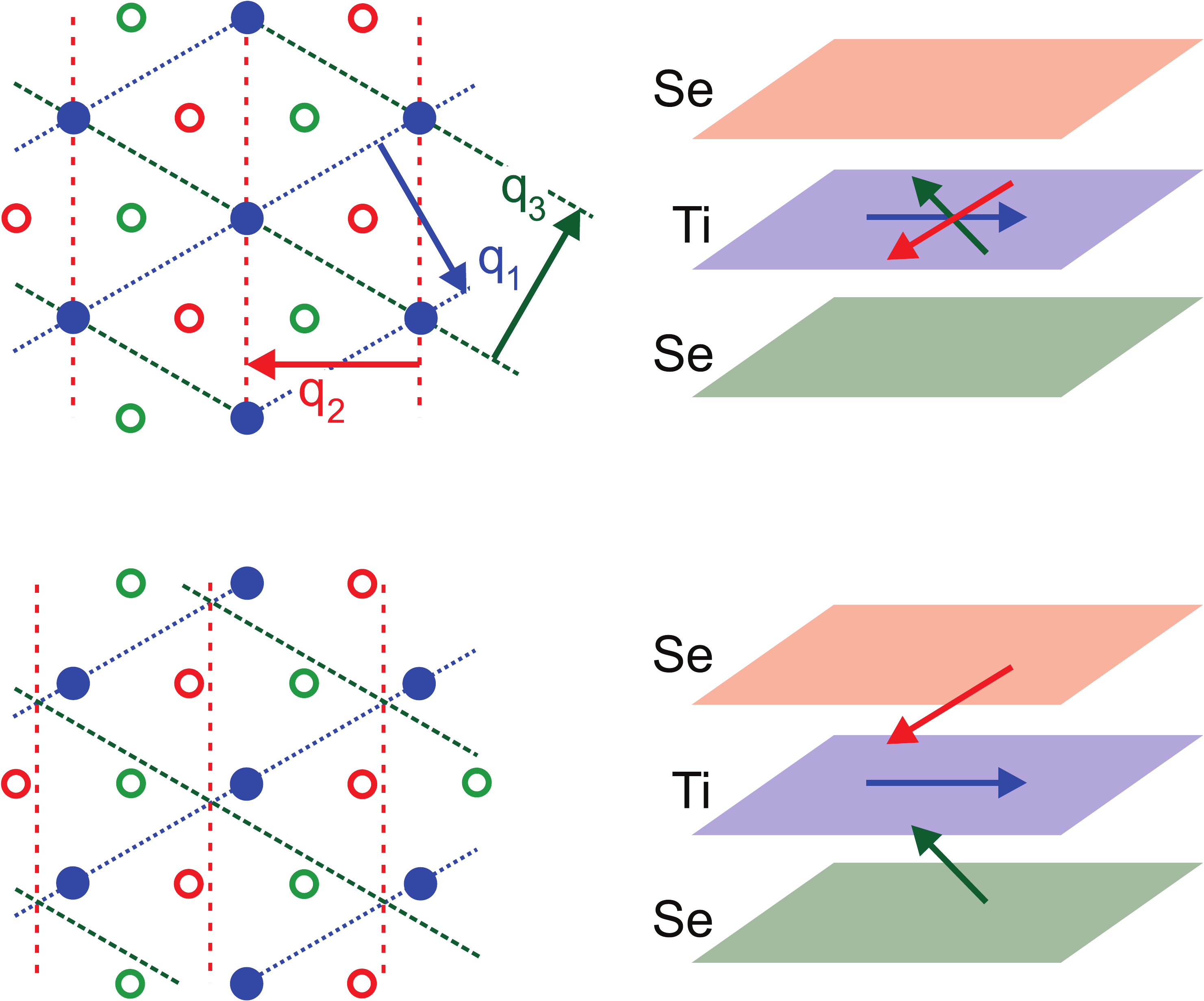}}}
\caption{(Color online) Schematic comparison of the non-chiral (top) and chiral (bottom) CDW phases of TiSe$_2$. The diagrams on the left show a projection of the atomic structure onto the crystallographic $ab$ plane. Ti ions are drawn as solid blue circles, while Se ions in the top and bottom layer are open red (light) and green (dark) circles respectively. Dashed lines indicate the extrema of each of the three CDW components. The arrows in the three-dimensional representations of the stacked Ti and Se layers on the right indicate the directions and the positions along the the crystallographic $c$-axis of the maximum amplitude displacements.}
\label{chiral}
\end{figure}

Evaluating the integrals of equation \eqref{F} yields:
\begin{align}
F &= \frac{3}{2} a_0^{\phantom 2} \psi_0^2 + \frac{1}{2} a_1^{\phantom 2} \psi_0^2 \left( 1-\gamma \right) \sum_j \cos\left( 2 \varphi_j \right) \notag \\
&+ \frac{3}{8} \left( 15 c_0^{\phantom 2} + 8 d_0^{\phantom 2} \right) \psi_0^4 + \frac{3}{4} c_2^{\phantom 2} \psi_0^4 \sum_j \cos \left( 2\varphi_j - 2\varphi_{j+1} \right), \notag
\end{align}
where we retained only the leading Umklapp terms. In this expression, both the coupling of the individual CDW components to the lattice and the interaction between them come from Umklapp effects. This can be contrasted with the case of elemental Te or Se, where the interaction between orbital components is proportional to $a_0$ and thus directly due to the Coulomb energy \cite{tbp}. For TiSe$_2$, the CDW involves charge transfer within the unit cell only, so that the total charge of each unit cell is automatically zero, even though the total charge density on individual atoms may not be zero. The term proportional to $c_2$ is required to minimize the resulting on-site Coulomb energy within the unit cell.

Minimizing the integrated free energy with respect to the phase variables yields the solutions:
\begin{align}
\varphi_1 = \frac{\pi}{2}, \ \varphi_2 = - \varphi_3 = \pm \frac{1}{2} \cos^{-1}\left[ \frac{3 c_2^{\phantom 2} \psi_0^2 - 2 a_1^{\phantom 2} \left(1-\gamma\right)}{6 c_2^{\phantom 2} \psi_0^2} \right]. \notag
\label{phis}
\end{align}
These solutions represent the chiral phases of TiSe$_2$, with the sign of the last variable determining the handedness. If the argument of the inverse cosine is less than $-1$, the lowest energy solution instead becomes the 
non-chiral triple-$q$ mode given by $\varphi_1=\varphi_2=\varphi_3=\pi/2$. 
\begin{figure}[t]
\centerline{{\includegraphics[width=0.5 \columnwidth]{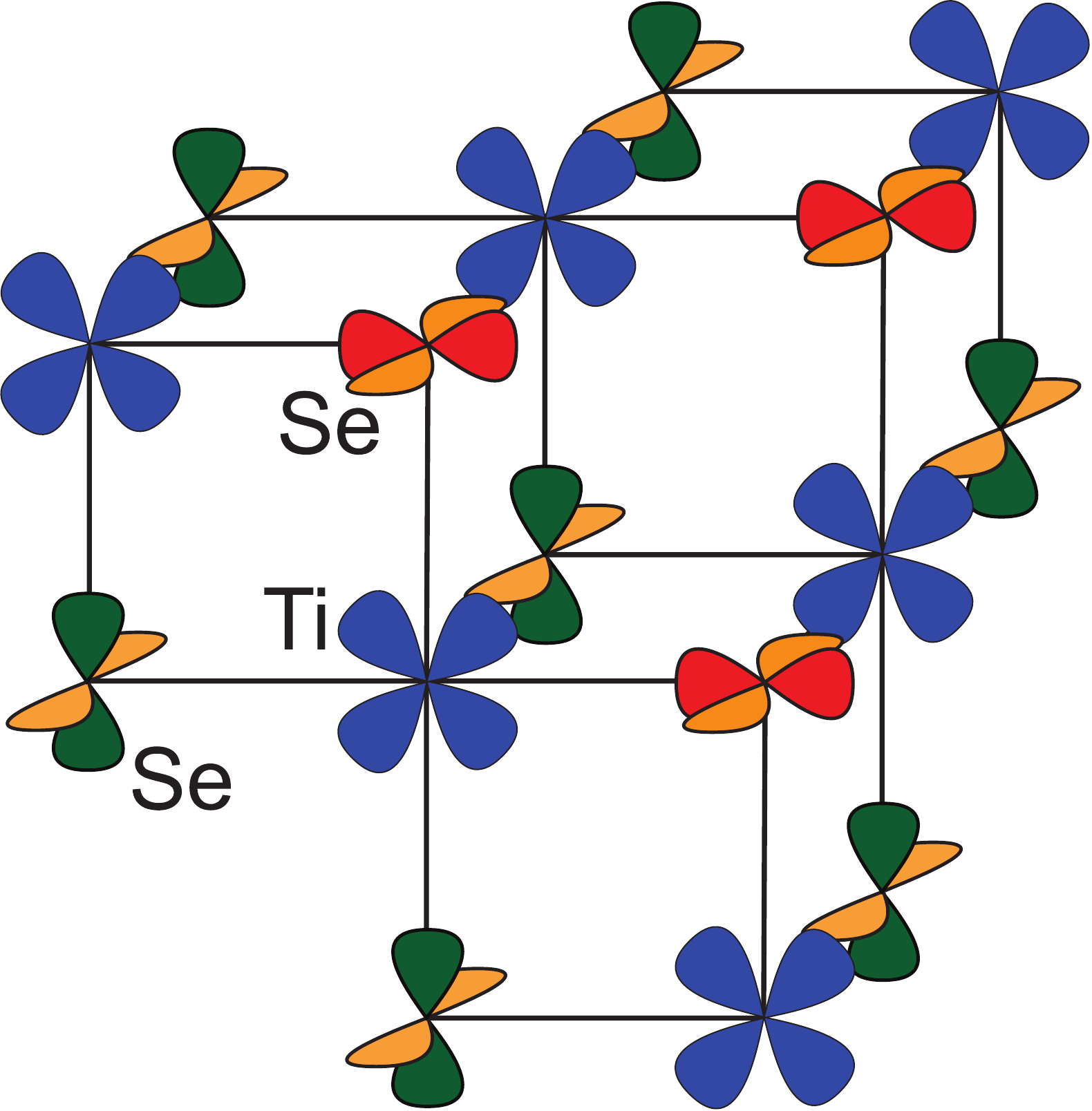}}}
\caption{(Color online) The orbital order corresponding to the chiral phase. The Se $p$ orbitals shown (two per atom) are the ones most depleted by the formation of the CDW, while the Ti $t_{2g}$ orbitals shown are the most occupied ones.}
\label{orbital}
\end{figure}
%

{\it The Chiral Phase.}--To visualize the chiral solution given above, we show on the right of Fig. \ref{chiral} the position of the maximum displacement wave amplitudes along the axis perpendicular to the Ti and Se sheets. In the usual, non-chiral triple-$q$ state all orbital components share the same phase, and the maximum atomic displacements all occur in the Ti layer. As we enter into the the chiral solution, the relative phase differences become nonzero, and they grow as the CDW amplitude continues to increase. The chirality of this phase can be clearly seen by considering the rotation of the dominant displacement direction as one traverses the material from the bottom Se layer to the top.

Alternatively, we can visualize the chiral phase by considering the occupation of the Ti $t_{2g}$ and Se $p$ orbitals. As the polarizations of the PDW components are directly linked to the orbitals involved, the dominant displacement direction in each atomic layer in the chiral phase corresponds to a maximum change in occupation for a single set of orbitals in that layer. Drawing the most affected orbitals in each layer then shows that the chiral phase is also an orbital ordered state, in which the plane containing these orbitals rotates between consecutive atomic layers (see Fig. \ref{orbital}). The same type of orbital order can also be identified in elemental Te and Se \cite{Fukutome:1984vs,tbp}. Indeed, since any chiral CDW will necessarily consist of components with different displacement wave polarizations, the formation of orbital order along with any chiral charge order is inevitable.

The condition which determines whether or not the chiral phase is a stable solution can be rewritten as a transition temperature separating the chiral phase from the usual triple-$q$ mode.  This yields a phase diagram where we first go from the uniform, high temperature state to the non-chiral triple-$q$ CDW at a critical temperature $T_{\text{CDW}}$, and only then enter into the chiral phase at a lower temperature $T_{\text{Chiral}}$:
\begin{align}
T_{\text{Chiral}} = T_{\text{CDW}} - \frac{a_1 (1-\gamma)}{9 \tilde{a} c_2} \left( 15 c_0 + 8 d_0 + 6 c_2 \right),
\end{align}
where we used the definition $a_0 = \tilde{a} (T-T_{\text{CDW}}) + a_1 (1-\gamma)$. In the absence of experimental estimates for the various Ginzburg-Landau parameters, we cannot make a quantitative prediction for $T_{\text{Chiral}}$. However, the chiral phase has been observed at $84$ K \cite{Ishioka:2010ez}, while the initial $T_{\text{CDW}}$ is well known to be $202$ K. The transition from the non-chiral to the chiral phase must lie somewhere in between. From the expression of the free energy above, it is straightforward to see that this must be a second order phase transition. If $T_{\text{Chiral}}$ turns out to differ substantially from $T_{\text{CDW}}$, higher order terms in the Ginzburg-Landau expansion may become relevant.

To observe the predicted phase transition between the two CDW phases of TiSe$_2$, various experimental techniques can be used. It was shown already that STM can directly image the chiral charge order \cite{Ishioka:2010ez}. Imaging the surface while the sample is being heated should thus yield a direct observation of the transition. Reflectometry may be used as an indirect probe of the chirality, since it shows the reduction of the usual three-fold symmetry of triple-$q$ modes to the two-fold symmetry of a chiral state \cite{Ishioka:2010ez}. This lowering of the lattice symmetry should also be visible in diffraction experiments by the emergence of previously forbidden peaks. The space group of TiSe$_2$ is lowered from $P\bar{3}m1$ to $P\bar{3}c1$ as the non-chiral CDW forms \cite{Anonymous:FHR9JE05}. The chiral phase breaks the symmetry down further, so that only a $P2$ space group remains. Alternatively, one could use the atomic pair distribution functions obtained in diffraction experiments to directly gauge the atomic positions. In the chiral phase the displacements in each layer are dominated by a single propagation vector, rather than the equal contributions from three vectors in the non-chiral phase. Finally, the orbital-specific nature of resonant inelastic X-ray scattering (RIXS) experiments may be used to probe the orbital order associated with the chiral phase.

The chiral order also allows for topological excitations. These come about because each layer in the TiSe$_2$ `sandwich' is dominated by a particular displacement direction. Different domains in the material may have varying handedness of the chirality as well as a different dominant CDW vector in the central Ti layer. A topological defect is formed when three domains with different CDW vectors meet at a point. This defect is somewhat similar to the vortices seen in phases with broken continuous U(1) symmetry (such as superconductors or superfluids), although in the present case there are only three distinct displacement directions.

{\it Conclusions.}--We have shown that the recently observed chirality in the charge ordered phase of TiSe$_2$ may be understood as a result of the interaction between three differently polarized displacement waves. The transversal polarizations of these waves are due to the relative orientations of the electronic orbitals involved in the charge transfer process that underlies the CDW formation. This central role played by the orbitals is reflected in the fact that the chiral charge order can equivalently be described as an orbital ordered state.

The part of the Ginzburg-Landau free energy that drives the formation of a chiral phase consists of generic Coulomb and Umklapp (electron-lattice coupling) terms. It is thus expected that a chiral CDW phase exists in many more materials which meet the basic prerequisites for displaying such a charge ordered state. To create a chiral pattern, a CDW must consist of at least three components with different polarizations of the associated lattice displacements \cite{footnote2}. Additionally, the shifts of the displacement waves resulting from the introduction of relative phases must have a common component which defines a propagation vector for the chiral state. Since the energetic avantage of the chiral state arises at least partly from Umklapp processes, it may also be expected that commensurate CDW materials are more prone to the formation of chiral order than incommensurate ones. One interesting material which does not have a CDW instability, but satisfies all other requirements, is 1T-TiTe$_2$. Since it has the same structure as TiSe$_2$, the series of compounds TiTe$_{2-x}$Se$_x$ should continuously interpolate between the two extremes, and thus have a chiral quantum critical point at some critical doping level $x_C$. 

The general nature of the terms in the free energy that drive the transition, in combination with the relatively short list of required material properties, suggests that there may well be a broad class of materials with a chiral charge and orbital ordered phase.

\subsection{Acknowledgements}
The author gratefully acknowledges insightful discussions with P.B. Littlewood and M.R. Norman, and support from the US DOE, Office of Science, under Contract No. DE-AC02-06CH11357.


\begin{thebibliography}{20}
\expandafter\ifx\csname natexlab\endcsname\relax\def\natexlab#1{#1}\fi
\expandafter\ifx\csname bibnamefont\endcsname\relax
  \def\bibnamefont#1{#1}\fi
\expandafter\ifx\csname bibfnamefont\endcsname\relax
  \def\bibfnamefont#1{#1}\fi
\expandafter\ifx\csname citenamefont\endcsname\relax
  \def\citenamefont#1{#1}\fi
\expandafter\ifx\csname url\endcsname\relax
  \def\url#1{\texttt{#1}}\fi
\expandafter\ifx\csname urlprefix\endcsname\relax\def\urlprefix{URL }\fi
\providecommand{\bibinfo}[2]{#2}
\providecommand{\eprint}[2][]{\url{#2}}

\bibitem[{\citenamefont{Ishioka et~al.}(2010)}]{Ishioka:2010ez}
\bibinfo{author}{\bibfnamefont{J.}~\bibnamefont{Ishioka}} \bibnamefont{et~al.},
  \bibinfo{journal}{Phys. Rev. Lett.} \textbf{\bibinfo{volume}{105}},
  \bibinfo{pages}{176401} (\bibinfo{year}{2010}).

\bibitem[{\citenamefont{van Wezel and Littlewood}(2010)}]{JvW:Physics11}
\bibinfo{author}{\bibfnamefont{J.}~\bibnamefont{van Wezel}} \bibnamefont{and}
  \bibinfo{author}{\bibfnamefont{P.}~\bibnamefont{Littlewood}},
  \bibinfo{journal}{Physics} \textbf{\bibinfo{volume}{3}}, \bibinfo{pages}{87}
  (\bibinfo{year}{2010}).

\bibitem[{\citenamefont{Overhauser}(1962)}]{Overhauser:1962gc}
\bibinfo{author}{\bibfnamefont{A.}~\bibnamefont{Overhauser}},
  \bibinfo{journal}{Phys. Rev.} \textbf{\bibinfo{volume}{128}},
  \bibinfo{pages}{1437} (\bibinfo{year}{1962}).

\bibitem[{\citenamefont{Littlewood}(1983)}]{Littlewood:1983fg}
\bibinfo{author}{\bibfnamefont{P.~B.} \bibnamefont{Littlewood}},
  \bibinfo{journal}{Crit. Rev. Solid State Mater. Sci.}
  \textbf{\bibinfo{volume}{11}}, \bibinfo{pages}{229} (\bibinfo{year}{1983}).

\bibitem[{\citenamefont{Mih{\'a}ly and
  J{\'a}nossy}(1984)}]{1984PhRvB..30.3530M}
\bibinfo{author}{\bibfnamefont{L.}~\bibnamefont{Mih{\'a}ly}} \bibnamefont{and}
  \bibinfo{author}{\bibfnamefont{A.}~\bibnamefont{J{\'a}nossy}},
  \bibinfo{journal}{Phys. Rev. B} \textbf{\bibinfo{volume}{30}},
  \bibinfo{pages}{3530} (\bibinfo{year}{1984}).

\bibitem[{\citenamefont{Gr{\"u}ner}(1988)}]{Gruner:1988wl}
\bibinfo{author}{\bibfnamefont{G.}~\bibnamefont{Gr{\"u}ner}},
  \bibinfo{journal}{Rev. Mod. Phys.} \textbf{\bibinfo{volume}{60}},
  \bibinfo{pages}{1129} (\bibinfo{year}{1988}).

\bibitem[{\citenamefont{Fukutome}(1984)}]{Fukutome:1984vs}
\bibinfo{author}{\bibfnamefont{H.}~\bibnamefont{Fukutome}},
  \bibinfo{journal}{Prog. Theor. Phys.} \textbf{\bibinfo{volume}{71}},
  \bibinfo{pages}{1} (\bibinfo{year}{1984}).

\bibitem[{tbp()}]{tbp}
\bibinfo{note}{A comparison between the microscopic theories of TiSe$_2$ and
  elemental Te and Se will be published separately.}

\bibitem[{\citenamefont{Wilson}(1977)}]{1977SSCom..22..551W}
\bibinfo{author}{\bibfnamefont{J.~A.} \bibnamefont{Wilson}},
  \bibinfo{journal}{Sol. State Comm.} \textbf{\bibinfo{volume}{22}},
  \bibinfo{pages}{551} (\bibinfo{year}{1977}).

\bibitem[{\citenamefont{Hughes}(1977)}]{Hughes:1977tg}
\bibinfo{author}{\bibfnamefont{H.~P.} \bibnamefont{Hughes}},
  \bibinfo{journal}{J. Phys. C} \textbf{\bibinfo{volume}{10}},
  \bibinfo{pages}{L319} (\bibinfo{year}{1977}).

\bibitem[{\citenamefont{Whangbo and Canadell}(1992)}]{Whangbo:1992vu}
\bibinfo{author}{\bibfnamefont{M.~H.} \bibnamefont{Whangbo}} \bibnamefont{and}
  \bibinfo{author}{\bibfnamefont{E.}~\bibnamefont{Canadell}},
  \bibinfo{journal}{J. Am. Chem. Soc.} \textbf{\bibinfo{volume}{114}},
  \bibinfo{pages}{9587} (\bibinfo{year}{1992}).

\bibitem[{\citenamefont{Cercellier et~al.}(2007)}]{Cercellier:2007tr}
\bibinfo{author}{\bibfnamefont{H.}~\bibnamefont{Cercellier}}
  \bibnamefont{et~al.}, \bibinfo{journal}{Phys. Rev. Lett.}
  \textbf{\bibinfo{volume}{99}}, \bibinfo{pages}{146403}
  (\bibinfo{year}{2007}).

\bibitem[{\citenamefont{van Wezel
  et~al.}(2010{\natexlab{a}})}]{vanWezel:2010uo}
\bibinfo{author}{\bibfnamefont{J.}~\bibnamefont{van Wezel}}
  \bibnamefont{et~al.}, \bibinfo{journal}{Europhys. Lett.}
  (\bibinfo{year}{2010}{\natexlab{a}}).

\bibitem[{\citenamefont{van Wezel
  et~al.}(2010{\natexlab{b}})}]{vanWezel:2010uc}
\bibinfo{author}{\bibfnamefont{J.}~\bibnamefont{van Wezel}}
  \bibnamefont{et~al.}, \bibinfo{journal}{Phys. Rev. B}
  \textbf{\bibinfo{volume}{81}}, \bibinfo{pages}{165109}
  (\bibinfo{year}{2010}{\natexlab{b}}).

\bibitem[{\citenamefont{Calandra and Mauri}(2011)}]{Calandra:2011bo}
\bibinfo{author}{\bibfnamefont{M.}~\bibnamefont{Calandra}} \bibnamefont{and}
  \bibinfo{author}{\bibfnamefont{F.}~\bibnamefont{Mauri}},
  \bibinfo{journal}{Phys. Rev. Lett.} \textbf{\bibinfo{volume}{106}}
  (\bibinfo{year}{2011}).

\bibitem[{\citenamefont{Rossnagel}(2011)}]{Rossnagel:2011jj}
\bibinfo{author}{\bibfnamefont{K.}~\bibnamefont{Rossnagel}},
  \bibinfo{journal}{J. Phys.: Cond. Matt.} \textbf{\bibinfo{volume}{23}},
  \bibinfo{pages}{213001} (\bibinfo{year}{2011}).

\bibitem[{\citenamefont{van Wezel
  et~al.}(2010{\natexlab{c}})}]{vanWezel:2010vq}
\bibinfo{author}{\bibfnamefont{J.}~\bibnamefont{van Wezel}}
  \bibnamefont{et~al.}, \bibinfo{journal}{Phys. Status Solidi}
  (\bibinfo{year}{2010}{\natexlab{c}}).

\bibitem[{\citenamefont{McMillan}(1975)}]{McMillan:1975wg}
\bibinfo{author}{\bibfnamefont{W.~L.} \bibnamefont{McMillan}},
  \bibinfo{journal}{Phys. Rev. B} \textbf{\bibinfo{volume}{12}},
  \bibinfo{pages}{1187} (\bibinfo{year}{1975}).

\bibitem[{\citenamefont{Wilson}(1978)}]{Anonymous:FHR9JE05}
\bibinfo{author}{\bibfnamefont{J.~A.} \bibnamefont{Wilson}},
  \bibinfo{journal}{Phys. Rev. B}  (\bibinfo{year}{1978}).

\bibitem[{foo()}]{footnote2}
\bibinfo{note}{Notice that we don't require
  out-of-plane components for the CDW vectors or the polarizations (cf.  \cite{Ishioka:2010ez}).}

\end{thebibliography}
\end{document}